# Results of direct dark matter detection with CDEX experiment at CJPL

**Hao Ma, Ze She, Zhongzhi Liu, Litao Yang, Qian Yue, Zhi Zeng, Tao Xue for the CDEX Collaboration**

Key Laboratory of Particle and Radiation Imaging (Ministry of Education) and Department of Engineering Physics, Tsinghua University, Beijing 100084, China

E-mail: mahao@tsinghua.edu.cn.

**Abstract**. The China Dark Matter Experiment (CDEX), located at the China Jinping Underground Laboratory (CJPL) whose overburden is about 2400 m rock, aims at direct searches of light Weakly Interacting Massive Particles (WIMPs). A single-element 994-gram p-type point contact (PPC) germanium detector (CDEX-1B) runs inside a solid passive shielding system. To achieve lower background, a prototype 10 kg PPC germanium detector array (CDEX-10), consisting of three detector strings with three germanium crystals each, is directly immersed in the liquid nitrogen. With the energy threshold of 160 eV, the limits on WIMP-nucleus scattering are set by energy spectra and annual modulation analysis, respectively. Incorporating Migdal effect, the data of CDEX-1B are re-analyzed to search sub-GeV WIMPs. Finally, the future plan of CDEX experiment in CJPL-II is introduced.

## 1. Introduction to CDEX

Various astronomical and cosmological evidences indicate the existence of dark matter, which contributes about one quarter of the energy density of the Universe. As the most desirable candidates of the dark matter, weakly interacting massive particles (WIMPs) could interact with nuclei of normal matter via elastic scattering, the recoil energy of which can be detected in extreme low background experiments at deep underground laboratories [1]. China Jinping Underground Laboratory (CJPL), located in Sichuan Province southwest of China, is an ideal site for dark matter searches with a rock overburden of 2400 meters [2]. Thanks to low energy threshold and good energy resolution, p-type point contact (PPC) germanium detectors, sensitive to sub-keV recoil energy, were used to search light WIMPs with masses 1 GeV/$c^2$ to 10 GeV/$c^2$ [3,4]. The China Dark Matter Experiment (CDEX) aims at direct detection of light WIMPs using PPC germanium detectors and started to run at CJPL in 2010 [5–7]. In this paper, results from CDEX-1B with a single-element 994-gram PPC germanium detector and CDEX-10 consisting of three detector strings will be reported. The future plan of CDEX experiment with several-hundred-kg PPC germanium detectors will also be presented.

## 2. Status and prospects of CDEX

Based on the first generation of CDEX (CDEX-1A) experiment, CDEX-1B improved the front-end electronics and structure materials close to the germanium crystal to achieve lower energy threshold and intrinsic background level. The single-element PPC detector was surrounded by combined shielding structure made of oxygen-free high-conductivity copper, borated polyethylene, and lead. In addition, the detector and copper bricks were placed in an acrylic box with a nitrogen gas flush to mitigate radon. The entire setup was installed in a room, with one-meter-thick polyethylene floor, roof and walls, at CJPL [8]. Based on 737.1 kg-day exposure with an analysis threshold of 160 eVee ("eVee" represents electron equivalent energy), first results from CDEX-1B improved over our previous limits on spin-independent (SI) and spin-dependent (SD) interacting cross-sections and extended the low reach of light WIMPs mass of 2 GeV/$c^2$. The new SD parameter space was probed with masses less than 4 GeV/$c^2$ [8].

The time dependence of WIMP-nucleon scattering signals, i.e. annual modulation effect, was also used to analyze the long-term data of CDEX-1B experiment. The data with a 1107.5 kg-day exposure were selected for annual modulation (AM) analysis with an energy threshold of 250 eVee [9]. For astrophysical-model-independent analysis, no positive results were found according to the best-fit of individual modulation amplitudes. Moreover, the limits on SI WIMP-nucleon interacting cross-sections were derived using standard spherical isothermal galactic halo model [10]. The results, shown in figure 1, provided more stringent bounds excluding 90% confidence level (C. L.) allowed regions implied by the DAMA/LIBRA phase 1 [11] and CoGeNT [12] at more than 99.99% and 98% C. L., respectively.

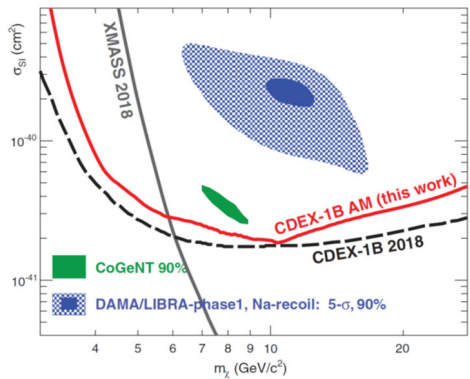

**Figure 1.** Limits at 90% C.L. from CDEX-1B AM-analysis (red) on spin-independent WIMP-nucleon cross section [9].

The conventional and simplified treatment of WIMP-nucleon scattering is that all the kinetic energy is transferred from WIMPs to nuclear recoil via elastic scattering. There is, however, finite probability that high-energy electrons are ejected via inelastic WIMP-nucleon scattering processes. The process, i.e. Migdal effect, was studied in the context of WIMP direct detection [13]. When nuclear recoils are not observable below the energy threshold of 160 eVee of typical CDEX PPC germanium detectors, the Migdal effect can still produce above-threshold signals and hence open the sensitivity windows to sub-GeV WIMPs. Analysis on time-integrated (TI) spectral and AM effects on CDEX-1B data are performed, with 731.7 kg-day exposure and 160 eVee threshold for TI analysis, and 1107.5 kg-day exposure and 250 eVee threshold for AM analysis. The sensitive windows in WIMP masses ($m_\chi$) are expanded by an order of magnitude to lower DM masses by incorporating Migdal effect. As shown in figure 2, new limits on SI cross-sections at 90% C. L. are derived as $2\times10^{-32} \sim 7\times10^{-35}$ cm$^2$ for TI analysis at $m_\chi \sim$50–180 MeV/$c^2$, and $3\times10^{-32} \sim 9\times10^{-38}$ cm$^2$ for AM analysis at $m_\chi \sim$75 MeV/$c^2$–3.0 GeV/$c^2$ [14].

Toward a future ton-scale DM experiment, the second generation CDEX experiment with a total detector mass of about 10 kg, called CDEX-10, has used three triple-element PPC germanium detector strings immersed in liquid nitrogen. The liquid nitrogen acts as cooling medium for germanium arrays and passive shielding material against ambient radioactivity simultaneously. The first 102.8 kg-day data from one of CDEX-10 PPC detectors were analyzed with an energy threshold of 160 eVee and achieved the background level of 2 counts/keV/kg/d between 2 and 4 keV. The improved limits of $8\times10^{-42}$ and $3\times10^{-36}$ cm$^2$ at a 90% C.L. on SI and SD WIMP-nucleon cross sections, respectively, at a $m_\chi$ of 5 GeV/$c^2$ are achieved [15].

The long-term physical goal of CDEX project is a ton-scale germanium experiment (CDEX-1T) searching for dark matter and $^{76}$Ge neutrinoless double beta decay process as well. The experimental setup will deploy germanium detector strings in a ~1725 m$^3$ liquid nitrogen tank, 16 meters in diameter and 18 meters in height. The tank is located in a pit with a diameter of 18 meters in and a height of 18 meters in the Hall C of the expansion project of CJPL (CJPL-II) [2]. The tank has been set up and will start commissioning in 2020. The next step of CDEX experiment (CDEX-100) is to adopt 100 kg germanium detector array to enlarge target mass and study background characteristics inside liquid nitrogen environment. In addition, the advanced study of future LEGEND1000 experiment [16] will also be carried out based on CDEX-100 system in parallel compared with LEGEND200 using liquid argon as active veto to be operated in Gran Sasso Underground Laboratory. The comparative study on background features is necessary to decide experimental technology roadmap of LEGEND1000.

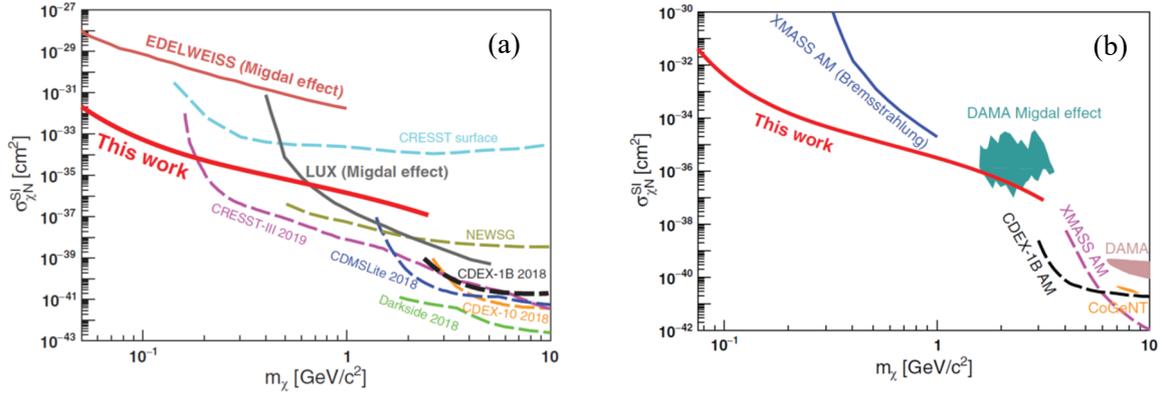

**Figure 2.** Upper limits at 90% C.L. on SI cross-sections derived in TI (a) and AM (b) analysis using the CDEX-1B experiment data [14].

## 3. Summary

The recent results of dark matter detection from CDEX-1B and CDEX-10 experiments were presented in this paper. Toward lower threshold and background level, CDEX-100 with a large liquid nitrogen tank was proposed as the next step of CDEX experiment and as the advanced study on background features for future LEGEND1000 neutrinoless double beta decay experiment.

**Acknowledgement**
This work is supported by the National Key Research and Development Program of China (No. 2017YFA0402201), the National Natural Science Foundation of China (Nos. 11675088, 11725522), and Tsinghua University Initiative Scientific Research Program (No. 20197050007).